\documentclass[%
 reprint,
 amsmath,amssymb,
 aps,
]{revtex4-2}

\usepackage{graphicx}
\usepackage{dcolumn}
\usepackage{bm}
\usepackage{xcolor}

\usepackage{braket} 
\usepackage{caption}
\usepackage{algorithm}
\usepackage{multirow}
\usepackage{algpseudocode}

\DeclareCaptionFormat{myformat}{#3}
\begin{document}

\preprint{APS/123-QED}

\title{Can flocking aid the path planning of microswimmers in turbulent flows?}

\author{Akanksha Gupta}
 \email{guptaakanksha17@gmail.com}
 \affiliation{
Department of Physics, Indian Institute of Science, Bangalore - 560012, India 
and
\\ Maulana Azad National Institute of Technology (MANIT), Bhopal-462003, M.P,  India 
}
\author{Jaya Kumar Alageshan}%
\email{jayaka@iisc.ac.in}
\author{Kolluru Venkata Kiran}%
\email{kiran8147@gmail.com}
\author{Rahul Pandit}
 \email{rahul@iisc.ac.in}
\affiliation{%
Department of Physics, Indian Institute of Science, Bangalore - 560012,India 
}%


\date{\today}

\begin{abstract}
We show that flocking of microswimmers in a turbulent flow can enhance the efficacy of reinforcement-learning-based path-planning of microswimmers in turbulent flows. In particular, we develop a machine-learning strategy that incorporates  Vicsek-model-type flocking in microswimmer assemblies in a statistically homogeneous and isotropic turbulent flow in two dimensions (2D). We build on the adversarial-reinforcement-learning of Ref.~\cite{alageshan2020machine} for non-interacting microswimmers in turbulent flows. Such microswimmers aim to move optimally from an initial position to a target. We demonstrate that our flocking-aided version of the adversarial-reinforcement-learning strategy of Ref.~\cite{alageshan2020machine} can be superior to earlier microswimmer path-planning strategies.

\end{abstract}

\maketitle


\noindent 

\section{Introduction}

Microswimmers, such as bacteria, algae, and microbots, are ubiquitous, and they play crucial roles in various biological and engineering processes, e.g., nutrient cycling, water-quality maintenance, disease transmission, and targeted drug delivery~\cite{deegan1997capillary,zaichik2008particles,schmidt2020engineering,mills1999targeted,hassanzadeh2019significance,bae2011targeted}. It is important, therefore, to understand how they navigate in different fluid flows. If the flow is turbulent, the optimal path planning of such microswimmers is especially difficult, because they are buffeted by the flow and  they can get trapped in eddies. This path planning has been addressed recently by bringing together methods from fluid mechanics and reinforcement learning~\cite{reddy2016learning,reddy2018glider,colabrese2017flow,alageshan2020machine,biferale2019zermelo,calascibetta2023taming}. Adversarial-reinforcement-learning strategies have been shown to help the path-planning of microswimmers~\cite{alageshan2020machine,biferale2019zermelo}. We demonstrate that flocking-aided path planning can, in some cases, outperform all microswimmer path-planning strategies that have been tried hitherto. 

Transport in response to stimuli, in general termed {\it taxis}~\cite{barrows2011animal}, is observed in a variety of systems ranging from cells and micro-organisms~\cite{dusenbery2009living} to birds, animals, and fish. Examples include chemotaxis~\cite{Wadhams2004Nature}, phototaxis~\cite{Monthiller:PRL2022}, and gravitaxis~\cite{Borge2014Nature,colabrese2017flow}. Interactions between the moving entities in these systems can lead to complex collective phenomena, e.g., flocking in schools of fish or the murmuration of starlings~\cite{marchetti2013hydrodynamics,cavagna2010scale}. The Vicsek model~\cite{vicsek1995novel} is  commonly used to study flocking phenomena. We explore how and when such flocking-aided path planning can optimise the movement of microswimmers in turbulent flows.

 The trajectories of microswimmers in turbulent flows are complex and difficult to control.  
 Therefore, the development of machine-learning strategies for the path-planning of such microswimmers 
 is a grand challenge that is of fundamental importance and which has implications for practical applications in targeted drug delivery and the locomotion of microbots~\cite{jang2019targeted,schmidt2020engineering,zou2024adaptive}. 
We develop a new machine-learning strategy that (a) incorporates Vicsek-model-type flocking in microswimmer assemblies in a turbulent flow [by generalising the study of Ref.~\cite{baggaley2015model} for such flocking microswimmers in a Taylor-Green vortical flow] and (b) builds on our work~\cite{alageshan2020machine}, on non-interacting microswimmers in turbulent flows, which uses adversarial reinforcement learning to optimise their path planning. Such microswimmers aim to move optimally from an initial position to a target. Similar studies have been carried out for the gravitaxis~\cite{colabrese2017flow} of microswimmers and their path planning via actor-critic reinforcement learning in two-dimensional (2D) turbulent flows~\cite{biferale2019zermelo}. For the gravitaxis of microswimmers, it has been shown~\cite{monthiller2022surfing} that surfers, which follow velocity gradients in the flow, can outperform microswimmers that use the path-planning strategy of Refs.~\cite{alageshan2020machine,colabrese2017flow}. We demonstrate that our flocking-aided version of the adversarial-reinforcement-learning strategy of Ref.~\cite{alageshan2020machine} can be superior to earlier microswimmer path-planning strategies.

\begin{figure*}[!ht]
	\includegraphics[width=17cm,height=6cm]{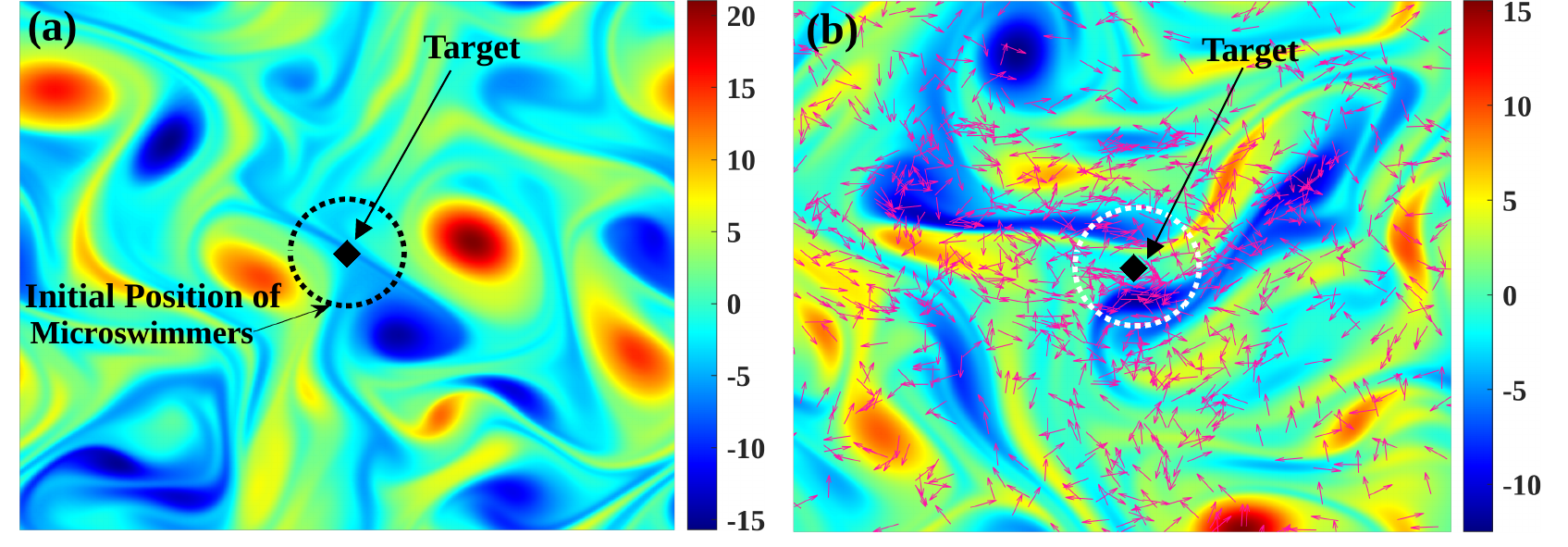}
	\caption{ (a) A pseudocolor plot of the vorticity field $\omega$ at the initial time; the target lies at the centre of the dotted circle, on which the microswimmers
 are distributed at random at the initial time. (b) A pseudocolor plot of $\omega$, at a representative time, with superimposed positions of some microflockers, whose swimming directions are indicated by red arrows (the length of each arrow is proportional to the speed of the microflocker). Each microflocker tries to align along the mean swimming direction of all microflockers in its neighborhood, a small circle of radius $R_0$ (we choose the illustrative value  $R_0=0.2\approx 0.03L$, where $L$ is the system size).}
	\label{fig:State_Color1}
\end{figure*}

\begin{figure*}[!ht]
	\includegraphics[width=18cm,height=17cm]{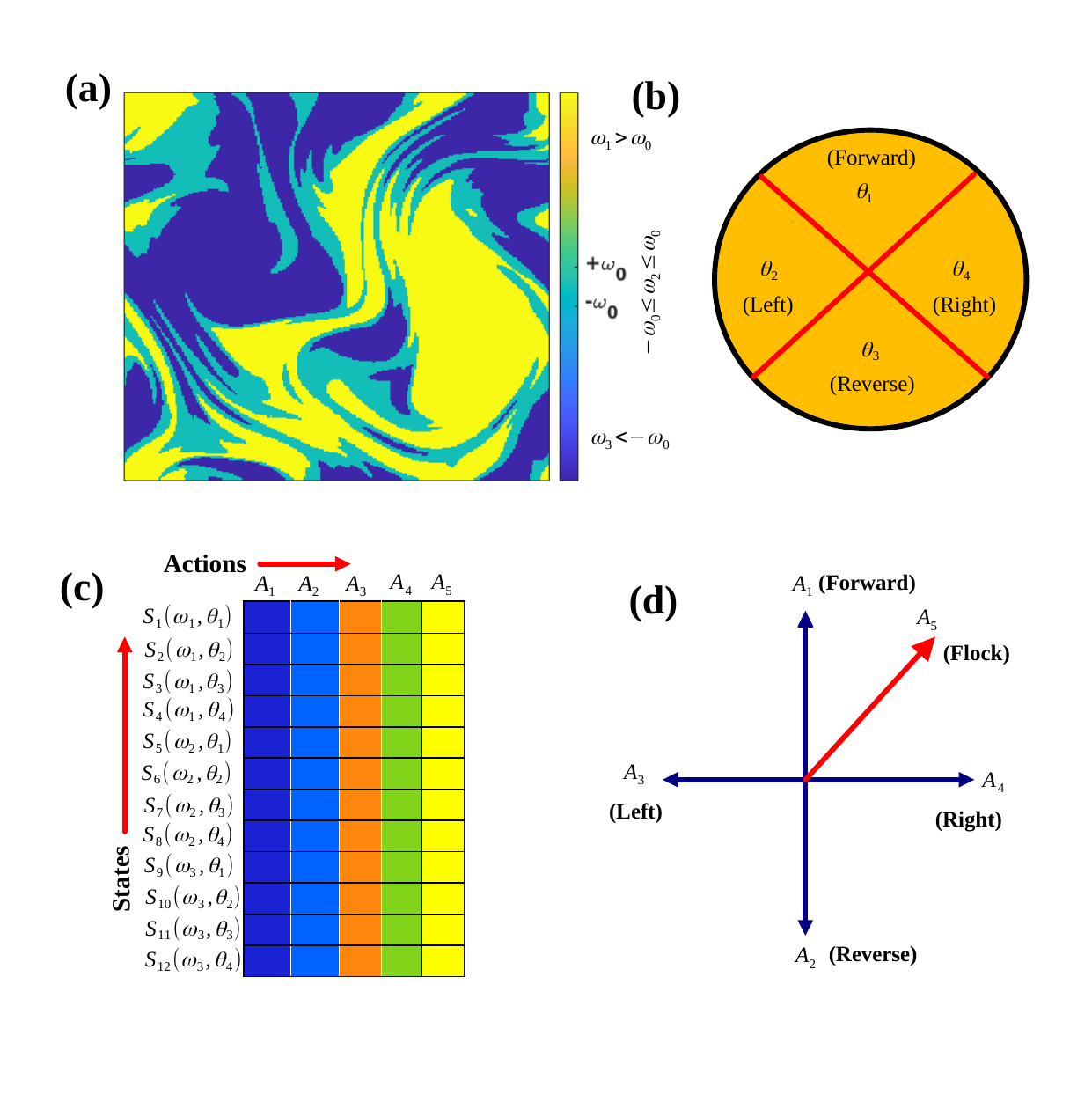}
	\caption{(a)  Discretization of the vorticity field  $\omega$ into 3 states,
	         namely, $\omega_1 \equiv \omega>\omega_{0}$, $\omega_2 \equiv -\omega_{0} \le \omega \le \omega_{0}$,
	         and $\omega_3 \equiv \omega < -\omega_{0}$;  (b) discretization of $\theta$: ($\theta_1 \equiv -\pi/4 \leq \theta < \pi/4$, 
	          $\theta_2 \equiv \pi/4\leq \theta < 3\pi/4$,  $\theta_3\equiv  3\pi/4\leq \theta <5\pi/4$,
	         $\theta_4\equiv  -3\pi/4 \leq \theta < -\pi/4$); (c) schematic diagram of our $\mathcal{Q}$ matrix with states in rows and actions in columns; (d) the set of actions $\{\mathbf{A}_{1},\mathbf{A}_{2},\mathbf{A}_{3},\mathbf{A}_{4},\mathbf{A}_{5}\}\equiv\{\hat{T},-\hat{T},\hat{T}_{\perp},-\hat{T}_{\perp},\arctan(\langle\hat{\mathbf{p}_j}\rangle_{n})\}$ (see text).}
	\label{fig:State_Color2}
\end{figure*}

The specific reinforcement-learning strategy we use is an adversarial $\mathcal{Q}$-learning method that is illustrated by the schematic diagrams in Figs.~\ref{fig:State_Color1} and ~\ref{fig:State_Color2}. We compare the following three types of microswimmers: (a) na\"ive swimmers ($NS$), (b) smart swimmers ($SS$), and (c) smart flockers ($SF$). The latter two, $SS$ and $SF$, are accompanied by a secondary microswimmer, which helps us to implement our adversarial strategy. Na\"ive swimmers reorient their direction towards the target, at every instant. Both $SS$ and $SF$ go beyond the $NS$ strategy by implementing an adversarial-$\mathcal{Q}$-learning algorithm~\cite{alageshan2020machine}. Smart flockers improve on the $SS$ strategy by interacting explicitly with their neighbors in a manner that mimics flocking.
 
We start with microswimmers distributed randomly on a circle; the target lies at the center of this circle [Fig.~\ref{fig:State_Color1}(a)].  In addition to their self-propulsion, all microswimmers are buffeted by a velocity field, which we obtain from a simulation of statistically steady, homogeneous, and isotropic turbulence in the two-dimensional (2D) Navier-Stokes equation.

Let $\hat{\mathbf{T}}\equiv(\mathbf{X}^{T}-\mathbf{X}(t))/\vert\mathbf{X}^{T}-\mathbf{X}(t)\vert$ be the unit vector from a microswimmer to the target, where $\mathbf{X}^{T}$ and $\mathbf{X}$ are the position vectors of the fixed target and a microswimmer, respectively. To develop a tractable framework for $\mathcal{Q}$ learning, we use the following discretization: (1) we use three states for the fluid vorticity $\omega$, at the microswimmer’s location, that are labelled by $\mathcal{S}_{\omega}\equiv\{\omega_{1},\omega_{2},\omega_{3}\}$; (2) for the angle $\theta\equiv \cos^{-1}\big(\hat{\mathbf{T}}\cdot\hat{\mathbf{p}}\big)$ between $\hat{\mathbf{T}}$ and $\hat{\mathbf{p}}$, we use four ranges $\mathcal{S}_{\theta}\equiv\{\theta_{1},\theta_{2},\theta_{3},\theta_{4}\}$ [see Fig.~\ref{fig:State_Color2}], where  $\hat{\mathbf{p}}$ is the swimming direction.  The final states are the $12$ elements [see  the $12$
   rows in Fig.~\ref{fig:State_Color2} (d)] of the set$~\mathcal{S}$, the Cartesian product of $\mathcal{S}_{\omega}$ and $\mathcal{S}_{\theta}$. 

  We denote actions generically by $\mathcal{A}$. For smart swimmers $a_{SS} \in \{\mathbf{A}_1,\mathbf{A}_2,\mathbf{A}_3,\mathbf{A}_4 \} (\equiv\mathcal{A}_{SS})$ [Fig.~\ref{fig:State_Color2} (d)].  We use one more direction as a possible action in the case of smart flockers: $a_{SF} \in \{\mathbf{A}_1,\mathbf{A}_2,\mathbf{A}_3,\mathbf{A}_4, \mathbf{A}_5\} (\equiv\mathcal{A}_{SF})$ [Fig.~\ref{fig:State_Color2} (d)]. Here, $\{\mathbf{A}_{1},\mathbf{A}_{2},\mathbf{A}_{3},\mathbf{A}_{4}\}\equiv\{\hat{\mathbf{T}},-\hat{\mathbf{T}},\hat{\mathbf{T}}_{\perp},-\hat{\mathbf{T}}_{\perp}\}$; and $\mathbf{A}_{5}\equiv\arctan(\langle\hat{\mathbf{p}}_{j}\rangle_{n})$, where $\langle \cdot \rangle_{n}$ denotes the sum over the neighborhood of the microswimmer, i.e., the sum over all the $j$ microswimmers that lie within a distance $R_0$ of the microswimmer under consideration, i.e., $\vert\mathbf{X}-\mathbf{X}_{j}\vert<R_{0}$, by virtue of which $\mathbf{A}_{5}$ incorporates a \textit{flocking action}, 
  as in the Vicsek model~\cite{vicsek1995novel} 
  where neighboring particles tend to point in the same flocking direction.
  The $12\times 4$ (or $12\times 5$) elements of the $\mathcal{Q}$ matrix [Fig.~\ref{fig:State_Color2} (d)] are given by 
  $\mathcal{Q}_{i,j}:(\mathfrak{S}_{i},A_{j})\to\mathbb{R}$, where $\mathfrak{S}_{i}\in\mathcal{S}$ and $A_{j}\in\mathcal{A}$ [see Sec.~\ref{Sec:model}
  for the Bellman equation for $\mathcal{Q}$ and the rewards that we use for the $SS$ and $SF$ cases].

The remainder of this paper is organised as follows: In Sec.~\ref{Sec:model} we define our model and describe the numerical methods we use. In Sec.~\ref{Sec:result} we present our results. Section~\ref{Sec:Conclusion} contains a discussion of the significance of our results. In the Appendix we present (a) the path-planning flowchart for the microswimmers in our model and (b) the pseudocode for our program.  

\section{Models and Numerical methods}
\label{Sec:model}
\subsection{The flow}

For the background flow, we consider statistically homogeneous isotropic turbulent flow in  a periodic square domain, with side $2 \pi$. This low-Mach-number fluid flow, with velocity $\mathbf{u}$, satisfies the incompressible Navier-Stokes equation which is
\begin{eqnarray}
	\partial_t\mathbf{\omega}+(\mathbf{u}.\nabla)\mathbf{\omega} = \nu \nabla^2 \mathbf{\omega}
	    -\alpha \omega + F_{\omega}\,; \;\;\;\; \nabla . \mathbf{u}=0 \,;
     \label{eq:NS}
\end{eqnarray}
here, $\nu$ and $\alpha$ are the fluid kinematic viscosity and coefficient of friction, respectively, the vorticity $\omega=\nabla \times \mathbf{u}$, the forcing term $F_{\omega}$ injects energy at large spatial scales and leads to statistically steady homogeneous and isotropic turbulence. We have verified that our results do not depend sensitively on the type of forcing by using either Kolmogorov forcing~\cite{Perlekar_PRL2011} or random forcing~\cite{alageshan2020machine}. For our direct numerical simulation (DNS) of Eq.~\eqref{eq:NS} we use a pseudospectral method, with a spatial resolution of $256\times 256$ collocation points, the $2/3$ dealiasing method for the removal of aliasing errors~\cite{orszag1971elimination,canuto1988ta}, and a second-order integrating-factor Runge-Kutta scheme for time marching~\cite{cox2002exponential}. We choose the time step $\Delta t$ to be small enough to satisfy the Courant-Friedrichs-Lewy (CFL) condition. The parameters in our DNSs are given in Table.~\ref{tab:Parameters}. 

\subsection{Microswimmer dynamics}
\label{subsec:microd}
We introduce microswimmers into the above background turbulent flow; they are situated initially at randomly chosen points on the dashed circle [radius  $R_p=1$] in Fig.~\ref{fig:State_Color1}(a); the target is fixed at the center of this circle. The microswimmer size is much smaller than the dissipation scale of the flow, so the position and orientation of the swimmer are given by~\cite{alageshan2020machine}
\begin{eqnarray}
	\frac{d\mathbf{X}}{dt} &=& \mathbf{u}(\mathbf{X},t) + V_s \; \hat{\mathbf{p}}\,, \nonumber \\
	\frac{d\hat{\mathbf{p}}}{dt} &=& \frac{1}{2B} \left[\hat{\mathbf{o}}-(\hat{\mathbf{o}}
	     . \hat{\mathbf{p}}) \hat{\mathbf{p}} \right] + \frac{1}{2} \omega \times
	      \hat{\mathbf{p}}\,,
	\label{eq:swimmer}
\end{eqnarray}
where $\mathbf{X}$ is the position of the microswimmer at time $t$, $\hat{\mathbf{p}}$ is the swimming direction, $V_s \hat{\mathbf{p}}$ is the swimming velocity, $B$ is the time scale of alignment, and $\hat{\mathbf{o}}$ is the preferred direction of motion of a na\"ive swimmer (NS) that likes to point towards the target. To solve Eqs.~\eqref{eq:swimmer}, we use the second-order Runge-Kutta method for time marching. We evaluate the fluid velocity and fluid vorticity at the location of the swimmer, which can lie off the collocation grid, by using bi-linear interpolation. The swimmers are passive because they have no back reaction on the flow velocity. We define the following non-dimensional parameters for these swimmers: $\tilde{V}_s \equiv V_s/u_{rms}$, $\tilde{B} \equiv B/\tau_\Omega$, and $\tau_\Omega \equiv \omega^{-1}_{rms}$ the inverse root-mean-squared vorticity. We consider both non-interacting and interacting microswimmers; in the latter case the interactions are induced by flocking, as in the widely studied Vicsek flocking model~\cite{vicsek1995novel}, which can show collective behavior with minimal constituents. 

\subsection{Microswimmers with flocking interactions}
\label{subsec:vicsek}

The Vicsek model for flocking can be used for the coordinated motion of self-propelled microswimmers, which have a tendency for flocking, as follows:
We define
\begin{equation}
\theta_i(t+1) = \langle \theta_j(t) \rangle_{R_0} + \zeta_i(t)\,,
\label{eq:VicsekTheta}
\end{equation}
where $\theta_i$ is the angle between the velocity of the $i^{th}$ microswimmer and the horizontal $x$ axis of the simulation domain and $\langle \theta_j \rangle_{R_0}$ is the average over the angles of neighboring particles that lie within the interaction radius $R_0$; we choose the representative value 
 $R_0=0.2\approx 0.03L$, where $L$ is the system size. In the original Vicsek model~\cite{vicsek1995novel}, the second term on the right-hand side (RHS) is zero-mean white noise. In our study, we do not consider this noise, i.e., $\zeta_i(t) = 0$, because the turbulence in our model is the source of dynamically generated noise. Furthermore, equations for the particle positions in the original Vicsek model~\cite{vicsek1995novel} are replaced by Eqs.~\eqref{eq:swimmer} for the microswimmers in our model.


\subsection{Reinforcement-learning algorithms for microswimmers}
\label{subsec:adverq}

Reinforcement learning (RL) is a machine-learning algorithm that involves training an agent to make decisions in an environment; certain decisions 
are favoured by rewards~\cite{sutton2018reinforcement,beintema2020controlling,gustavsson2016preferential,colabrese2017flow,qiu2020swimming, Landin:scirobotics}. In our study, the agents are microswimmers; and the background environment is the 2D turbulent flow described above. We allow microswimmers to explore different states [Fig.~\ref{fig:State_Color2}] and then provide a reward if the microswimmers choose the best possible action as specified in detail below. Both smart microswimmers and smart microflockers learn to swim towards the target using the $\epsilon-$greedy and
$\mathcal{Q}-$learning algorithms that we discuss below in Subsections~\ref{subsubsec:egreedy} and \ref{subsubsec:qlearn}. 

\subsubsection{The $\epsilon$-greedy method}
\label{subsubsec:egreedy}
In this method~\cite{Watkins_1992} a probability distribution decides the control direction and balances exploration and exploitation in $\mathcal{Q}$-learning. For our problem we choose~\cite{alageshan2020machine} the probability distribution function (PDF)
\begin{eqnarray}
\label{eq:epsilon_greedy}
    \mathbf{P}[{\bf{\hat{o_i}}}(s_i)]&=&\frac{\epsilon_g}{N_a}+(1-\epsilon_g)\delta({\bf{\hat{o}_i}}(s_i)-{\bf{\hat{o}}}_{max})\,,\nonumber\\
     {\bf{\hat{o}}}_{max}&:=& \text{argmax}_{a \in \mathcal{A}}\mathcal{Q}(s,a)\,,
\end{eqnarray}
\noindent where $N_a$ is the number of actions, which is $4$ and $5$ for smart microswimmers and smart microflockers, respectively, and
$\delta(\cdot)$ is the Dirac delta function. In Eq.~\eqref{eq:epsilon_greedy}, $\epsilon_g$ is the probability of exploration, and the agents choose a random action (exploration) with probability $\epsilon_g$.

    \subsubsection{ $\mathcal{Q}$-learning scheme}
\label{subsubsec:qlearn}

We use Bellman's equation, which is a Markov-decision process, i.e., for a given $\mathcal{Q}$-matrix
\begin{eqnarray}
\mathcal{Q}(\bm s(t),\hat{\bm
    o}(t)) &\mapsto& (1-\lambda)\,\mathcal{Q}(\bm s(t),\hat{\bm o}(t)) 
                         \nonumber \\ 
                         +&\lambda&\,\left[ r(t)+ \gamma\,\max_{\hat{\bm a}} Q(\bm
  s(t+\Delta t),\hat{\bm a}) \right]\,,
  \label{eq:GenQ}
\end{eqnarray}
where $\lambda$ and $\gamma$ are learning parameters that need to be set [see Table~\ref{tab:Parameters}]. We must consider
two such matrices, namely, $\mathcal{Q}_{SS}$ and $\mathcal{Q}_{SF}$, with 
\begin{eqnarray}
     \hat{\mathbf{o}}_{SS}:= \text{argmax}_{a \in \mathcal{A}_{SS}} \; 
     \mathcal{Q}_{SS}(s,a_{SS})\;\; {\rm{or}} \nonumber  \\
     \hat{\mathbf{o}}_{SF} := \text{argmax}_{a \in \mathcal{A}_{SF}} \; \mathcal{Q}_{SF}(s,a_{SF})\,,
     \label{eq:ohats}
  \end{eqnarray}
in Eq.~\eqref{eq:swimmer}; we use $\hat{\mathbf{o}} = \hat{\mathbf{o}}_{SS}$
and $\hat{\mathbf{o}} = \hat{\mathbf{o}}_{SF}$ for smart microswimmers and smart microswimmers, respectively.
\begin{table}[!ht]
	\begin{tabular}{|c | c | c|}
	    \hline
			Symbol & Value & Description 
                \\
                \hline
                $\nu$ & $2\times10^{-3}$& kinematic viscosity\\
                \hline
                 $\alpha$ & $5\times10^{-2}$& coefficient of friction\\
                \hline
	     $\Delta t$& $ 5\times 10^{-4}$ & time step\\
                \hline
           $\gamma$ & $9.9\times10^{-1}$ & learning discount\\
           \hline
           $\lambda$& $5\times10^{-3}$& learning rate\\
           \hline
           $\omega_{rms}$ & $4$ & root mean squared vorticity\\
           \hline
            $u_{rms}$ & $1.6$ & root mean squared velocity\\
           \hline
           $R_{0}$ & $0.2$ & interaction radius\\
           \hline
           $R_{p}$ & $1.0$ & initial location circle radius\\
           \hline
           $r_{\delta}$ & $0.062$ & targets capture-radius\\
           \hline
           $L$ & $2\pi$ & System size\\
           \hline
           $\epsilon_{g}$ & $1\times10^{-3}$ & greedy parameter\\
           \hline
           $\omega_{0}$ & $2.0$ & vorticity value for discretization\\
           \hline
           $\mathcal{N}_p$ & $1024$ & number of particles\\
           \hline
           $N$ & $256$ & resolution\\
           \hline
	
	\end{tabular}
	\caption{List of non-dimensional parameter values.}
	\label{tab:Parameters}
\end{table}  
\subsubsection{Adversarial $\mathcal{Q}$ learning and Rewards}
\label{subsubsec:reward}

We build on the work of Ref.~\cite{alageshan2020machine}, which employs adversarial $\mathcal{Q}$ learning for every smart microswimmers $SS$, by using a similar learning scheme for every smart microflocker $SF$. In this scheme, we associate one na\"ive microswimmer $NS$, referred to as the \textit{slave}~\cite{alageshan2020machine}, per $SS$ and $SF$, referred to as the \textit{master}~\cite{alageshan2020machine}; we use these terms in the sense
of master-slave multi-agent reinforcement learning~\cite{kong2017revisiting}. When a master changes its state, the associated slave's position and direction are reinitialized to that of the master [see Ref.~\cite{alageshan2020machine} for details]. We calculate the reward functions for the masters from  positions, denoted generically by $\mathbf{X}$, as follows:
\begin{eqnarray}
    r_{SS}({\mathcal{S}}(t),\hat{\bm o}_{SS}(t))=|\mathbf{X}^s_{SS}(t)-\mathbf{X}^T|-|\mathbf{X}_{SS}(t)-\mathbf{X}^T|\,; \nonumber \\
     \tilde{r}_{SF}({\mathcal{S}}(t),\hat{\bm o}_{SF}(t))=|\mathbf{X}^s_{SF}(t)-\mathbf{X}^T|-|\mathbf{X}_{SF}(t)-\mathbf{X}^T|\,;
\end{eqnarray}
here, the superscripts $s$ and $T$ denote the slave and the target and the subscripts $SS$ and $SF$  stand for smart microswimmers and microflockers, respectively.
 In our simulations we monitor $N_{SS}(t/\tau_\Omega)$, $N_{SF}(t/\tau_\Omega)$, and $N_{NS}(t/\tau_\Omega)$ that denote, respectively, the numbers of smart microswimmers, smart micro flockers, and na\"ive swimmers that reach the target up until the nondimensionalised time $t/\tau_\Omega$, where $\tau_\Omega = \omega_{rms}^{-1}$, the inverse of the root-mean-square vorticity. Once any swimmer reaches the target, we reintroduce it into our simulation domain at a point that is chosen randomly 
 on the circumference of the dashed circle shown in Fig.~\ref{fig:State_Color1} (a). To encourage flocking action, we define the final reward
 for $SF$ to be

\begin{equation}
\label{eq:reward_SF}
     r_{SF}(t)= 
\begin{cases}
    2\tilde{r}_{SF},& \text{if } \tilde{r}_{SF}> 0 \, \& \, a_{SF}=\langle \hat{\mathbf{p}} \rangle\,, \\\
    \tilde{r}_{SF},              & \text{otherwise} \,.
\end{cases} 
\end{equation}

\section{Results}
\label{Sec:result}

\begin{figure}[!ht]
\begin{center}
   \includegraphics[width=7.5cm,height=17cm]{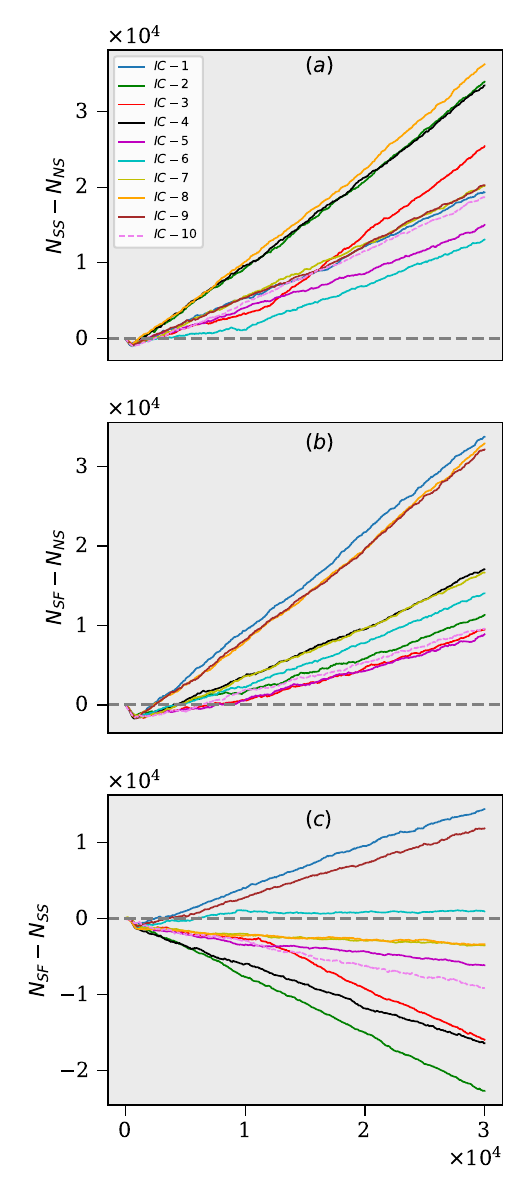}
  
   \caption{Plots versus the nondimensionalised time $t/\tau_\Omega$ of the differences (a) $(N_{SS}-N_{NS})$, (b) $(N_{SF}-N_{NS})$, and (c) $(N_{SF}-N_{SS})$, 
   where $N_{SS}$, $N_{SF}$, and $N_{NS}$ denote, respectively, the total numbers of smart microswimmers, smart microflockers, and na\"ive swimmers that have reached the target up until $t/\tau_\Omega$ (see text) for $\tilde{B}=2.0$ and $\tilde{V}_{s}=1.25$.}
               \label{fig:number_swimmers}
               \end{center}
\end{figure}

\begin{figure*}[!ht]
\begin{center}
\includegraphics[width=18.5cm,height=18cm]{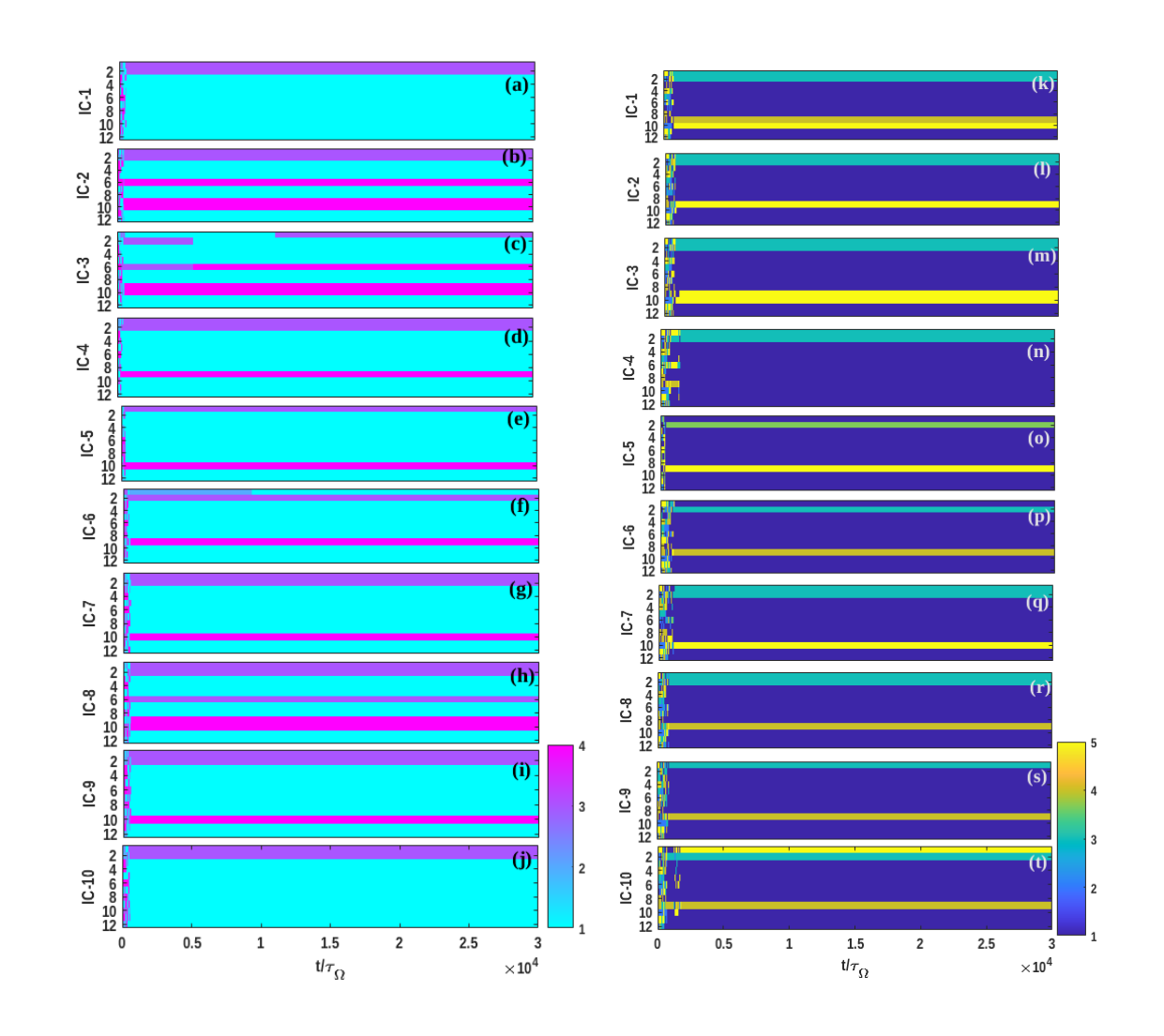}
 \caption{Plots versus the normalised time of the maximal value in a row of the $\mathcal{Q}$ matrix, for all $12$ states, for smart microswimmers (left panel) and smart microflockers (right panel) for $\tilde{B}=2.0$ and $\tilde{V}_{s}=1.25$. The colorbar shows the four (five) actions chosen in the left (right) panels by smart microswimmers (smart microflockers); we present plots for $10$ different initial conditions (a)-(j) in the left panel [(k)-(t) in the right panel]. The lemon-yellow band (right panel) shows when the flocking action is dominant.}
\label{fig:SF-SS}
  \end{center}
\end{figure*}

\begin{figure}[!ht]
\begin{center}
   \includegraphics[width=9cm,height=6cm]{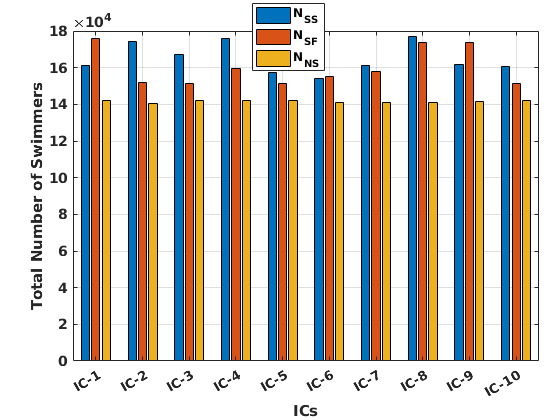}
  
   \caption{Bar charts of $N_{SS}$, $N_{SF}$, and $N_{NS}$ (see text and Fig.~\ref{fig:number_swimmers}) at the end of our simulation for the $10$ initial conditions IC1-IC10. 
           }
               \label{fig:T_number}
               \end{center}
\end{figure}

\begin{figure*}[!ht]
\begin{center}
   \includegraphics[width=18cm,height=6cm]{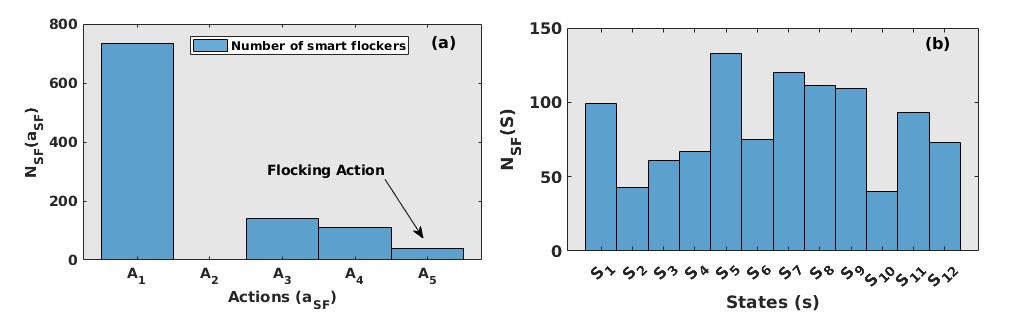}
\caption{Bar charts of $N_{SF}$ at the end of our simulation for initial condition IC-1 showing distributions (a) over actions the possible actions $a_{SF}$ and (b) the states $S_i$, $i= 1 \ldots 12$.}
\label{fig:hist}
\end{center}
\end{figure*}

 We have conducted calculations to determine the cumulative sum of microswimmers reaching the target at each time step.
Our principal results are given in Fig.~\ref{fig:number_swimmers} for ten distinct turbulent initial conditions (IC-1 to IC-10). The plots versus $t/\tau_\Omega$ of $(N_{SS}-N_{NS})$, $(N_{SF}-N_{NS})$, and $(N_{SF}-N_{SS})$ in Figs.~\ref{fig:number_swimmers}(a), (b), and (c), respectively, show that, as time increases, both $SS$ and $SF$ microswimmers benefit from their $\mathcal{Q}$-learning strategies and outperform their NS counterparts at large times. Figure~\ref{fig:number_swimmers}(c) shows clearly that there are three initial turbulent conditions (IC-1, IC-6, and IC-9) for which the smart microflockers $SF$ perform better than smart microswimmers $SS$. Note that the plots in Fig.~\ref{fig:number_swimmers} first dip and then rise; this illustrates the 
learning process that we can visualise as in Figs.~\ref{fig:SF-SS}(a) and (b) via plots versus the normalised time of the maximal value in a row of the $\mathcal{Q}$ matrix, for all $12$ states, for smart microswimmers and microflockers, respectively, for the illustrative parameter 
values $\tilde{B}=2.0$ and $\tilde{V}_{s}=1.25$. In Fig.~\ref{fig:T_number} we present bar charts of $N_{SS}$, $N_{SF}$, and $N_{NS}$ at the end of our simulation for the $10$ initial conditions IC1-IC10. Furthermore, the bar charts in Fig.~\ref{fig:hist}(a,b) present the distribution of microswimmers across five actions and twelve states, respectively, at the end of our simulation, specifically for the initial condition IC-3; in particular, Fig.~\ref{fig:hist}(a) clearly demonstrates that a significant number of microswimmers have chosen flocking, signifying the effectiveness of this action.



\begin{figure}[!ht]
   \includegraphics[width=8cm,height=16cm]{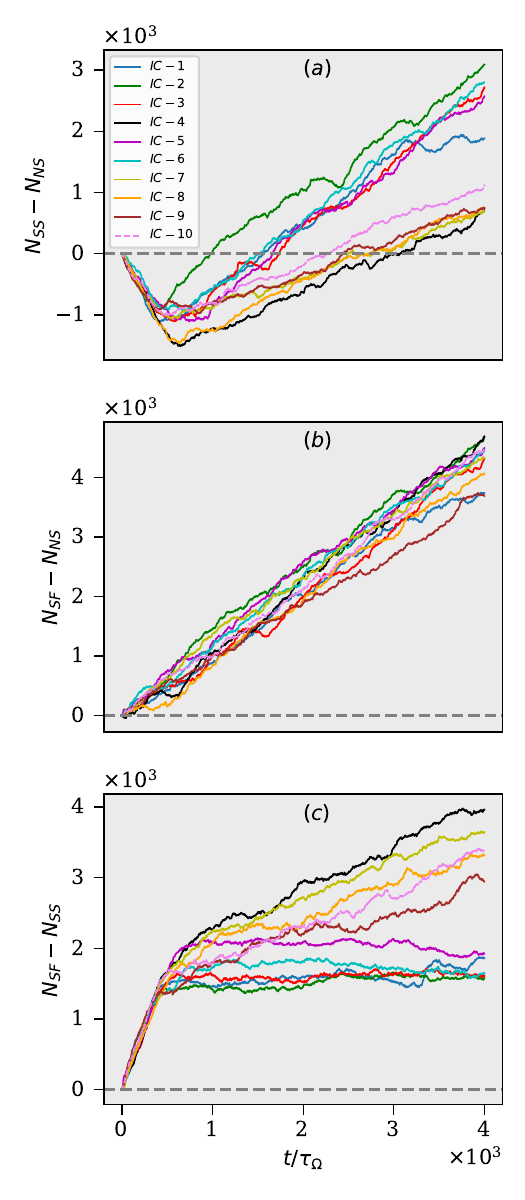}
\caption{ Plots versus the nondimensionalised time $t/\tau_\Omega$ of the differences (a) $N_{SS}-N_{NS}$, (b) $N_{SF}-N_{NS}$, and (c) $N_{SF}-N_{SS}$ for the 10 different initial conditions, 
   where $N_{SS}$, $N_{SF}$, and $N_{NS}$ denote, respectively, the total numbers of smart microswimmers, smart microflockers, and na\"ive swimmers that have reached the target up until $t/\tau_\Omega$ (see text), with \textit{transfer learning} of smart microflockers (see text), in which we use the optimised $\mathcal{Q}$ matrix of smart microflockers for $\tilde{B}=2.0$ and $\tilde{V}_{s}=1.25$.}
\label{fig:no_learning}
\end{figure}

\begin{figure*}[!ht]
   \includegraphics[width=17cm,height=17cm]{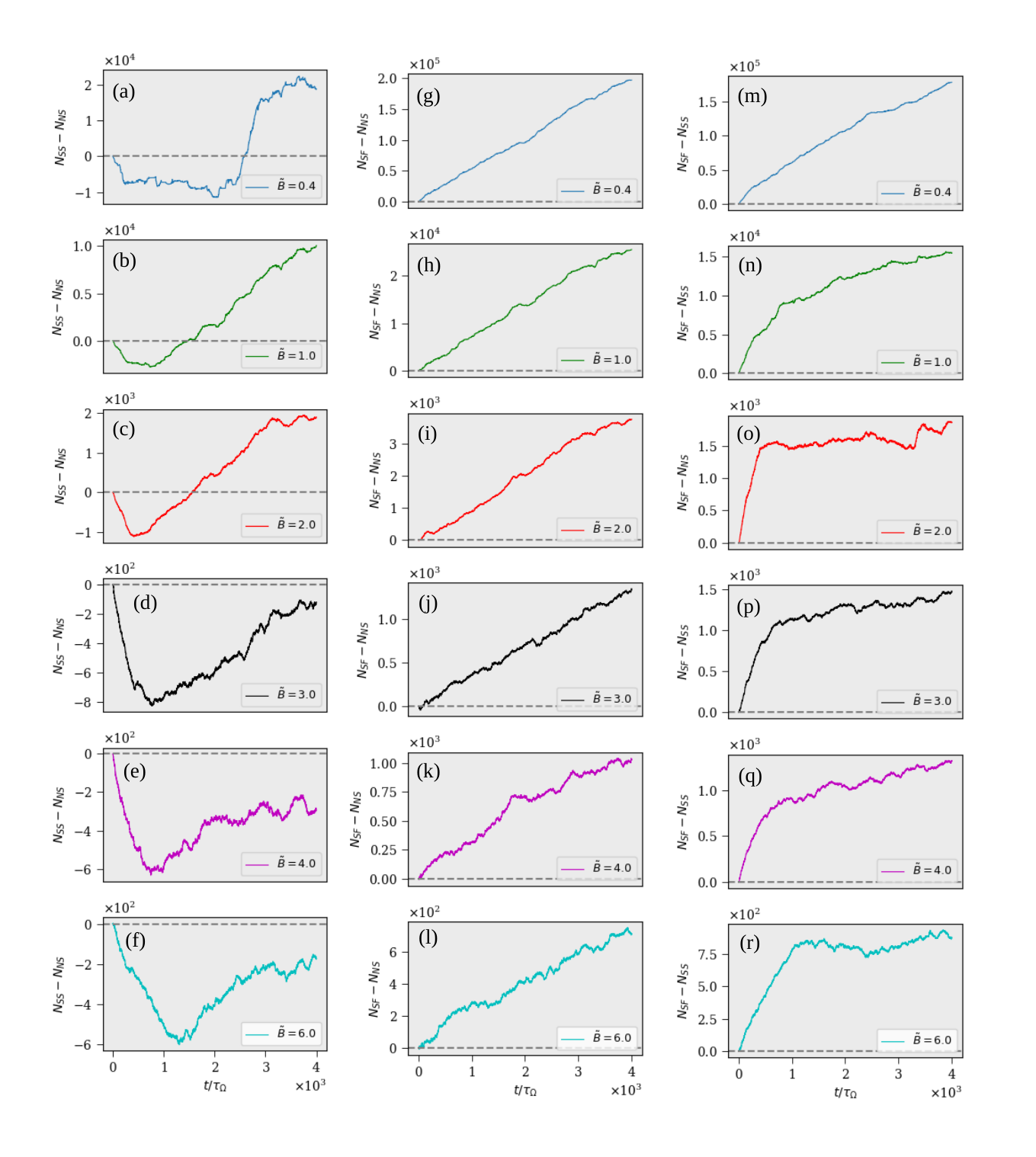}
\caption{Plots versus the nondimensionalised time $t/\tau_\Omega$ of the differences (a-f) $N_{SS}-N_{NS}$, (g-l) $N_{SF}-N_{NS}$, and (m-r) $N_{SF}-N_{SS}$, 
   where $N_{SS}$, $N_{SF}$, and $N_{NS}$ denote, respectively, the total numbers of smart microswimmers, smart microflockers, and na\"ive swimmers that have reached the target up until $t/\tau_\Omega$ (see text), with \textit{transfer learning} of smart microflockers (see text), in which we use the optimised $\mathcal{Q}$ matrix of smart microflockers for initial condition IC-1 for $\tilde{V_s}=1.25$ and various normalised values of $\tilde{B}=0.4,1.0,2.0,3.0, 4.0$, and $6.0$ .}
\label{fig:transfer_learning_B}
\end{figure*}

\begin{figure*}[!ht]
   \includegraphics[width=16cm,height=13cm]{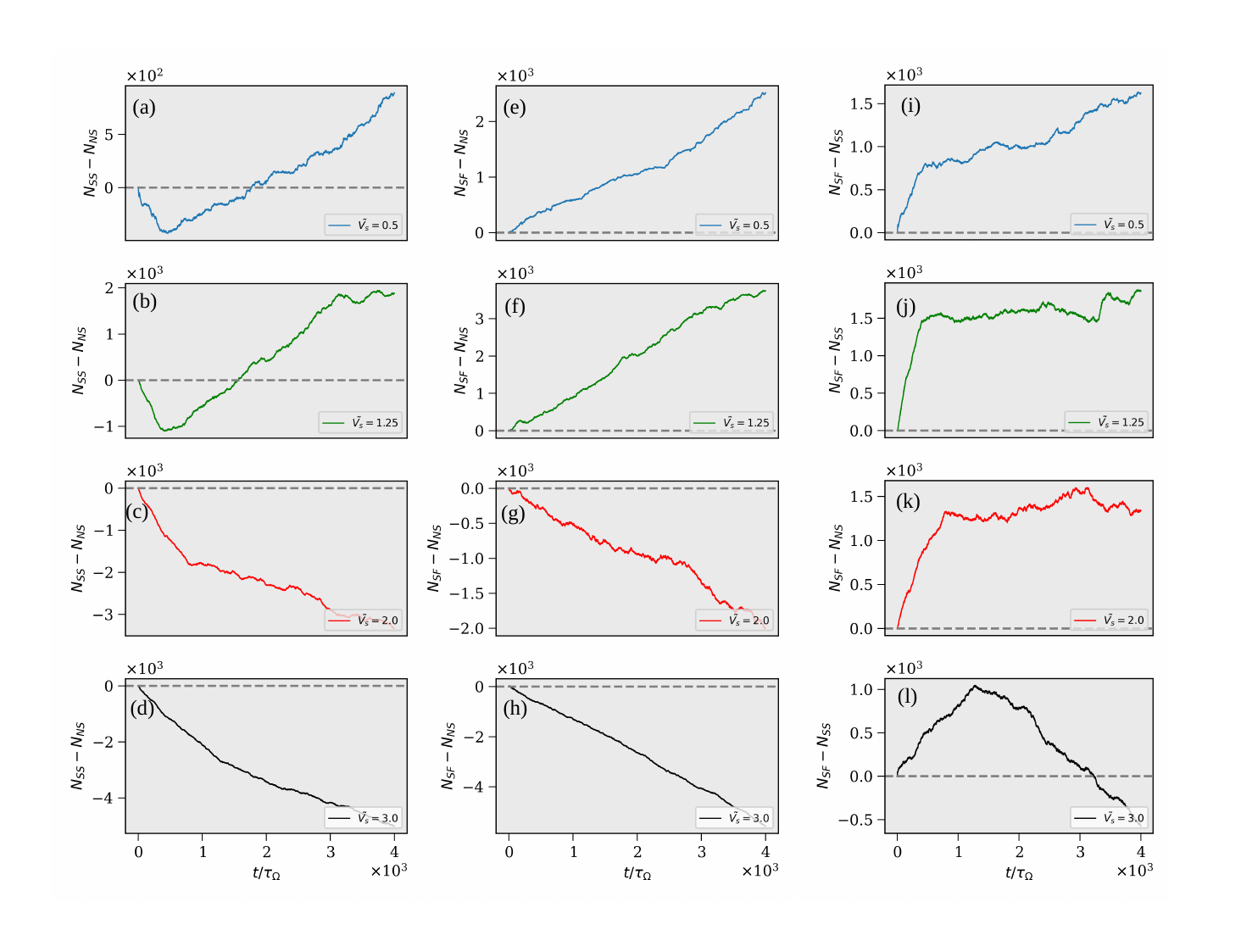}
\caption{Plots versus the nondimensionalised time $t/\tau_\Omega$ of the differences (a-d) $N_{SS}-N_{NS}$, (e-h) $N_{SF}-N_{NS}$, and (i-l) $N_{SF}-N_{SS}$, 
   where $N_{SS}$, $N_{SF}$, and $N_{NS}$ denote, respectively, the total numbers of smart microswimmers, smart microflockers, and na\"ive swimmers that have reached the target up until $t/\tau_\Omega$ (see text), with \textit{transfer learning} of smart microflockers (see text), in which we use the optimised $\mathcal{Q}$ matrix of smart microflockers for initial condition IC-1 for $\tilde{B}=2.0$ and various normalised values of $\tilde{V_s}=0.5,1.25,2.0,3.0$.    }
\label{fig:transfer_learning_Vs}
\end{figure*}

Transfer learning allows us to explore the knowledge acquired by previously trained agents (microswimmers in our case) to enhance the learning of new microswimmers for the same task but \textit{in different environments}. This transfer learning improves the performance of smart microswimmers by eliminating exploration. In particular, we select the $\mathcal{Q}$-matrix, which we obtain from the best-performing case [Fig.~\ref{fig:SF-SS} (k) with initial condition IC1], to evaluate the performance of trained microflockers after the learning process. 
With this  $\mathcal{Q}$-matrix and the initial condition IC-1, we obtain the most favourable outcome for smart microflockers which now surpass 
both na\"ive and smart microswimmers. In particular, we use the $\mathcal{Q}$-matrix from the final configuration of Fig.~\ref{fig:SF-SS} (k)
to carry out simulations for the ten different initial conditions IC1-IC10 [Fig.~\ref{fig:no_learning}] and different combinations of the parameters 
$\tilde{B_s}$ and $\tilde{V_s}$ [Figs.~\ref{fig:transfer_learning_B} and \ref{fig:transfer_learning_Vs}, respectively].

The plot of $(N_{SF}-N_{NS})$ versus $t/\tau_\Omega$ in Fig.~\ref{fig:no_learning}(a) shows a dip at
$t \simeq 1000\tau_\Omega$ followed by a rise, which means that smart microswimmers continue to learn until $t \simeq 1000\tau_\Omega$.  The plots in Figs.~\ref{fig:no_learning}(b) and (c) show that the number differences $(N_{SF}-N_{NS})$ and $(N_{SF}-N_{SS})$ increase as time passes; furthermore, Fig.~\ref{fig:no_learning}(c) shows that $N_{SF}>N_{SS}$ for the entire duration of our simulation.

The plots in Fig.~\ref{fig:transfer_learning_B}  versus the nondimensionalised time $t/\tau_\Omega$ of (a-f) $(N_{SS}-N_{NS})$, (g-l) $(N_{SF}-N_{NS})$, and (m-r) $(N_{SF}-N_{SS})$ show that, for $\tilde{V_s}=1.25$ and $\tilde{B} = 0.4\,, 1.0\,, 2.0\,, 3.0\,, 4.0\,, \rm{and}\, 6.0$, 
and  $\mathcal{Q}$-matrix and the initial condition IC-1 mentioned above, smart microflockers $SF$ continue to outperform the other microswimmers
[see, in particular, the last column of Fig.~\ref{fig:transfer_learning_B}]. Similar plots in Fig.~\ref{fig:transfer_learning_Vs}  versus the nondimensionalised time $t/\tau_\Omega$ of (a-d) $(N_{SS}-N_{NS})$, (e-h) $(N_{SF}-N_{NS})$, and (i-l) $(N_{SF}-N_{SS})$ show that, 
for $\tilde{B}=2.0$ and $\tilde{V_s}=0.5\,,1.25\,,2.0,$ and $3.0$, illustrate the following: smart microswimmers $SS$ outperform naive swimmers $NS$ when $\tilde{V_s} < u_{rms}$ [Fig.~\ref{fig:transfer_learning_Vs}(a) and (b)]; however, as the swimming speed increases, $SS$ do not exhibit superior performance relative to $NS$. By contrast, smart microswimmers $SF$ continue to outperform $SS$ [Fig.~\ref{fig:transfer_learning_Vs}(1-l)] except at the highest value of of the swimmer speed $\tilde{V_s}$, where the na\"ive strategy of $NS$ always wins.



\citet{Monthiller:PRL2022} have studied the gravitaxis of plankters in three dimensions. They have proposed a scheme that allows plankters, swimming at a constant speed, to move upward by choosing a swimming direction by following a surfing strategy that uses local flow gradients as follows:  The preferred direction in this surfing strategy is $\hat{\bf n}_{surf}={\bf n}_{surf}/\lvert {\bf n}_{surf} \rvert$, where ${\bf n}_{surf}=[\exp(\tau_{surf}({\bf\nabla u})]^{T}\cdot \hat{z}$, where $\tau_{surf}$ is a free parameter in their surfing strategy. Reference~\cite{Monthiller:PRL2022} then shows that plankters, which follow this strategy, achieve net vertical speeds that can be up to twice their swimming speed in a turbulent flow. 
\begin{figure}[!ht]
   \includegraphics[width=8cm,height=12cm]{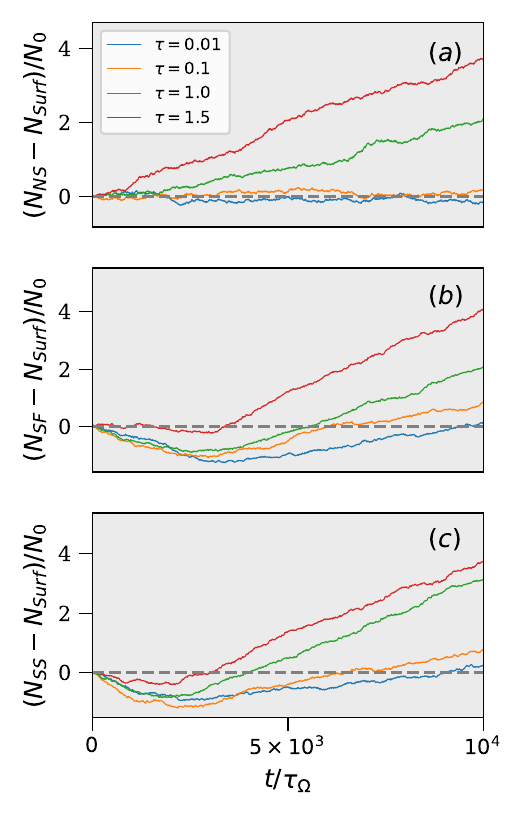}[h]
\caption{Plots versus the nondimensionalised time $t/\tau_\Omega$, for $\tilde{B}=5.0$ and $\tilde{V_s}=0.75$, of the differences (a) $(N_{NS}-N_{Surf})/N_0)$, (b)$(N_{SF}-N_{Surf})/N_0)$, and (c) $(N_{SS}-N_{Surf})/N_0)$, with $N_{Surf}$ the total numbers of surfers that reach the target up until $t/\tau_\Omega$ and $N_0$ is the number of microswimmers at $t/\tau_\Omega=0$. }
\label{fig:surfer}
\end{figure}
 We use the strategy of  Ref.~\cite{Monthiller:PRL2022} for microswimmers in the 2D turbulent flow that we consider; we refer to such swimmers as surfers. In our problem, we replace $\hat{z}$ by $\hat{\mathbf{T}}$. We then compare the total number of na\"ive swimmers $NS$, smart microswimmers $SS$, and smart microflockers $SF$ with surfers that have $\tau_{surf}=0.01\,, 0.1\,,  1.0$, and $1.5$ [in normalized unit $\tilde{\tau}=\tau_{surf}/\tau_\Omega=0.04\,, 0.4\,, 4.0$ and $6.0$, respectively]. The plots in Fig.~\ref{fig:surfer} show that, for high values of $\tilde{\tau}$, e.g., $\tilde{\tau}=4.0$ and  $6$, all microswimmers, $NS$, $SS$, and $SF$, perform better than surfers; however, for low values of $\tilde{\tau}$, e.g., $\tilde{\tau}=0.04$ and $0.4$, only na\"ive swimmers $NS$ perform as well as surfers. 
 
 Reference~\cite{Monthiller:PRL2022} estimates the optimal value $\tau_{surf}\simeq4\tau_\eta$ for $V_s=u_\eta$, where $\tau_\eta$ and $u_\eta$ are, respectively, the Kolmogorov time and velocity of the turbulent flow. In our study, which differs significantly from Ref.~\cite{Monthiller:PRL2022} in terms of spatial dimension and the control direction $\hat{\mathbf{o}}$, the swimmer speed $V_s$, the alignment time scale $B$, it is
 natural to choose a range $\tau_{surf} \in [10^{-2}-10^{0}]$. After training the smart microswimmers, we observe that the number of successful smart microswimmers that reach the target surpass the number surfers 
 that reach this goal, at  large values of $\tau_{surf}$ for $\tilde{B}=5.0$, and $\tilde{V}_s=0.75$. A detailed comparison 
 of our smart swimmers and flockers with surfers will be addressed in future work.
 

\section{Conclusions}
\label{Sec:Conclusion}

We have addressed the challenging problem of the path planning of microswimmers in turbulent flows by developing a machine-learning strategy that combines Vicsek-model-type flocking of such microswimmers with the adversarial $\mathcal{Q}$-learning method developed in Ref.~\cite{alageshan2020machine} for \textit{non-interacting} microswimmers in such flows. Note that flocking induces an effective interaction between microswimmers. For specificity, we consider microswimmers that aim to move optimally from an initial position to a target. We compare na\"ive swimmers $NS$ with smart swimmers $SS$, \`a la Ref.~\cite{alageshan2020machine}, and the smart flockers $SF$, which follow the algorithm given in Subsection~\ref{subsec:vicsek}. Our results show that $SF$ outperform $NS$ and $SS$ microswimmers for different values of $V_s$ and $B$ when the hyperparameters in Table~\ref{tab:Parameters} are optimised.
We find that, in certain ranges of the parameters, $V_s$ and $B$, and with the optimised hyperparameters in Table~\ref{tab:Parameters}, the smart flockers  $SF$ can outperform $NS$ and $SS$ microswimmers.
Our results are of relevance to microswimmers in a variety of experimental settings in which flocking is relevant~\cite{nitin_Nature14}.

\section*{Data and code availability}
Data from this study and the computer scripts can be obtained from the authors upon 
reasonable request.
  
\section*{Conflicts of Interest}
No conflicts of interest, financial or otherwise, are declared by the authors.

\section*{Author Contributions} 
JKA, RP, and AG planned the research; AG, JKA, and KVK carried out the calculations and analysed the numerical data; AG prepared the tables, figures, and the draft of the manuscript; AG, JKA, KVK, and RP then revised the manuscript in detail and approved the final version.

\section*{Funding}
We thank the Science and Engineering Research Board (SERB), BRNS, the National Supercomputing Mission (NSM), India, for support,  and the Supercomputer Education and Research Centre (IISc) for computational resources.  AG thanks DST-SERB for the NPDF fellowship (PDF/2021/001681).

\section*{Acknowledgments}
We thank Sriram Ramaswamy, Vasanth Kumar, and Nadia Bihari Pradhan for the valuable discussions.  


\appendix
\label{sec:appen}
\section{Pseudocode}
  \label{sec:algo_q_learning}
We present the pseudocode for calculating the reward after choosing the best possible action from the epsilon-greedy algorithm or the $\mathcal{Q}$-matrix.
\begin{figure}[!ht]

   \includegraphics[scale=0.5]{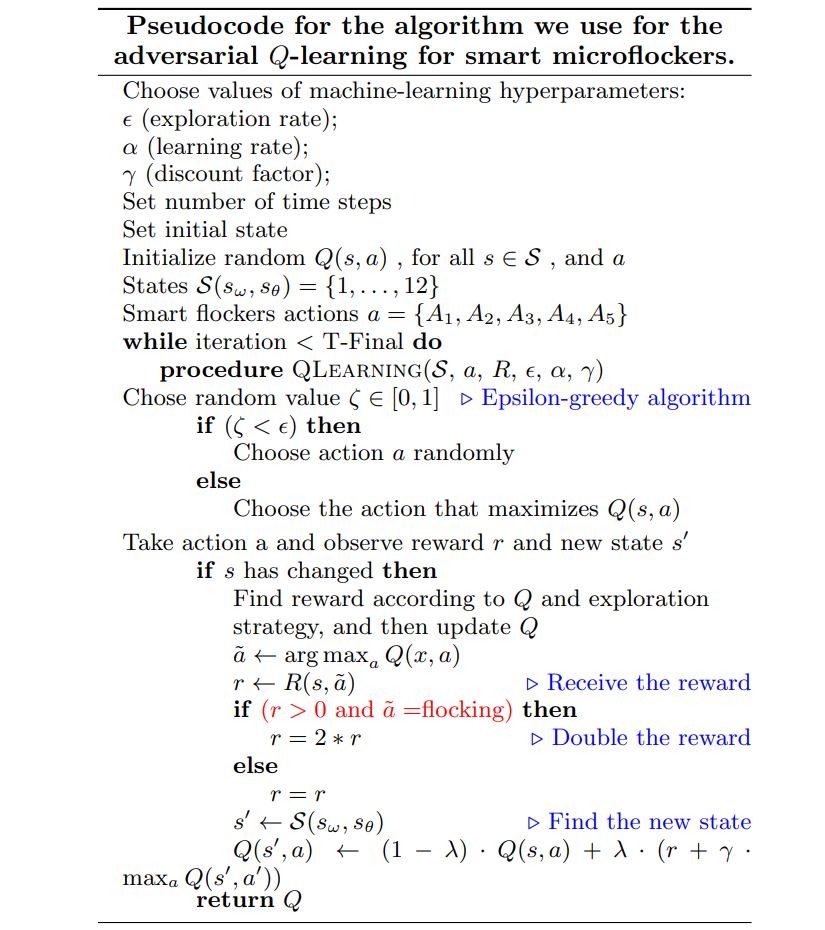}
  
            \label{fig:algorithm}
\end{figure}

\begin{figure}[!ht]

   \includegraphics[width=7cm,height=15cm]{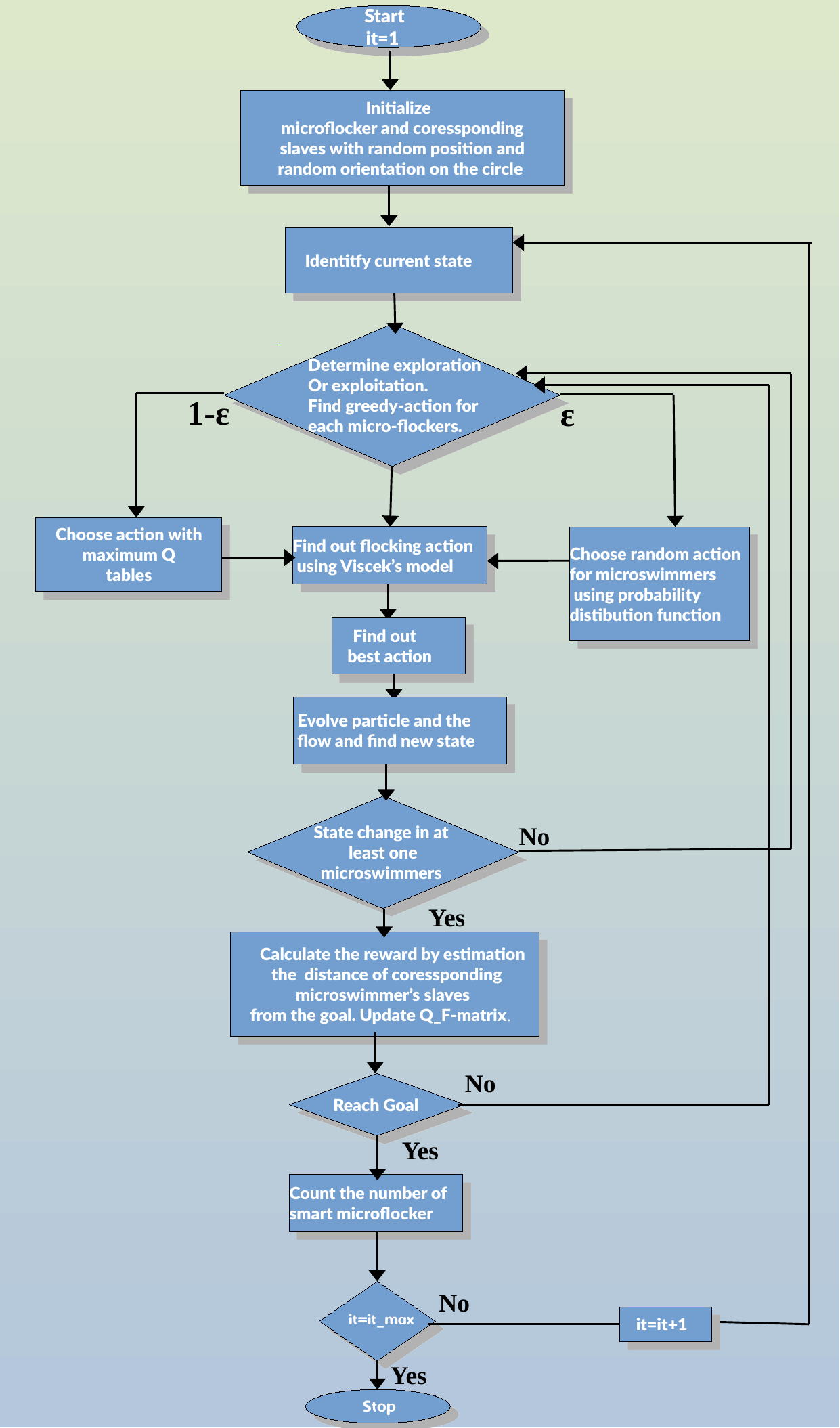}
   \caption{Flowchart with the sequence of processes involved in our
adversarial $\mathcal{Q}$-learning scheme for smart microflockers.
           }  
            \label{fig:flowchart}
\end{figure}

Fig.~\ref{fig:flowchart} demonstrates the sequence of processes involved in our
adversarial $\mathcal{Q}$-learning scheme for smart micro-flockers.

\end{document}